# An Automated Computational Pipeline for Generating Large-Scale Cohorts of Patient-Specific Ventricular Models in Electromechanical *In Silico* Trials

Ruben Doste, Julia Camps, Zhinuo Jenny Wang, Lucas Arantes Berg, Maxx Holmes, Hannah Smith, Marcel Beetz, Lei Li, Abhirup Banerjee, Vicente Grau, Blanca Rodriguez

## Aims

In recent years, human *in silico* trials have gained significant traction as a powerful approach to evaluate the effects of drugs, clinical interventions, and medical devices. *In silico* trials not only minimise patient risks but also reduce reliance on animal testing.

However, the implementation of *in silico* trials presents several time-consuming challenges. It requires the creation of large cohorts of virtual patients. Each virtual patient is described by their anatomy with a volumetric mesh and electrophysiological and mechanical dynamics through mathematical equations and parameters. Furthermore, simulated conditions need definition including stimulation protocols and therapy evaluation. For large virtual cohorts, this requires automatic and efficient pipelines for generation of corresponding files.

In this work, we present a computational pipeline to automatically create large virtual patient cohort files to conduct large-scale *in silico* trials through cardiac electromechanical simulations. The pipeline generates the files describing meshes, labels, and data required for the simulations directly from unprocessed surface meshes. We applied the pipeline to generate over 100 virtual patients from various datasets and performed simulations to demonstrate capacity to conduct *in silico* trials for virtual patients using verified and validated electrophysiology and electromechanics models for the context of use. The proposed pipeline is adaptable to accommodate different types of ventricular geometries and mesh processing tools, ensuring its versatility in handling diverse clinical datasets. By establishing an automated framework for large scale simulation studies as required for *in silico* trials and providing open-source code, our work aims to support scalable, personalised cardiac simulations in research and clinical applications.

## Introduction

The increasing availability of computational resources and credibility of multiscale modelling and simulation are supporting performance and reliability of large-scale studies using virtual patient cohorts for the evaluation of therapies and investigations of disease phenotypes. Thus, *in silico* trials, through simulations in large virtual patient cohorts, are becoming established as a powerful, ethical and cheap approach for testing the safety and efficacy of clinical interventions, new medical devices, and treatments (Dasí et al. 2024; Niederer et al. 2020; Pappalardo et al. 2019; Viceconti et al. 2021). Cardiac *in silico* trials use large cohorts of patients to enable the evaluation of various hypotheses and factors affecting human variability, such as inter-patient geometry, patient-specific physiology, comorbidities, and different stages or manifestations of the disease. This capability allows for the simulation and study of pathological mechanisms (Coleman et al. 2024; Strocchi et al. 2020), treatment strategies (Dasí et al. 2024; Doste et al. 2020; Roney et al. 2022; Yu et al. 2021), identification of new therapeutic targets (Margara et al. 2022), and evaluation of drug effects (Aguado-Sierra et al. 2022; Dasí et al. 2023; Krause, Zisowsky, and Dingemanse 2018). These simulations often incorporate "Digital Twins", where patient-specific geometries and data are integrated into the study (Corral-Acero et al. 2020; O'Hara et al. 2022; Roney et al. 2023). Additionally, *in silico* trials aim to reduce the need for animal experimentation, as previous work has demonstrated higher accuracy than animal models in predicting pro-arrhythmic cardiotoxicity (Passini et al. 2017).

Similar to clinical trials, *in silico* trials require a substantial number of patient-specific models and simulations to have a good coverage of population variability. This presents a challenge for organ-level studies, as generating patient-specific models is a highly time-consuming task. Consequently, establishing robust and efficient pipelines for generating and processing patient-specific models on a large scale is crucial (Niederer et al. 2020; Rodero et al. 2023). The availability of large databases, such as the UK Biobank (Sudlow et al. 2015), further enables the creation of virtual patient cohorts, provided that automated pipelines are applied.

In cardiology, recent advances in image segmentation algorithms (Banerjee et al. 2021; Beetz et al. 2023) and automatic meshing software (Neic et al. 2020) have facilitated cardiac geometry reconstruction. Currently, the main limitation for large-scale cohort creation lies in data preparation for simulations. Generating all the data required for the simulation such as fibre information, personalised electrical activation, cellular heterogeneities or geometry information (ventricular coordinates, fibrosis, remodelled tissue) necessitates a significant amount of manual intervention, thereby impeding efficient model creation. Often, the generation of this information requires the application of different algorithms, including remeshing, and using external software. This also implies the need to process the geometry to establish distinct initial conditions for each method's application. Examples of these algorithms include mesh processing tools (Fedele and Quarteroni 2021; Neic et al. 2020), cardiac coordinates generation (Bayer et al. 2018; Schuler et al. 2021), which facilitate the generation of coordinates to enable data transfer between different cardiac geometries, rule-based methods for myofiber orientation generation (Africa et al. 2023; Bayer et al. 2012; Doste et al. 2019; Piersanti et al. 2021), algorithms for personalisation to patient electrophysiology (Camps et al. 2021, 2024; Gillette, Matthias A.F. Gsell, et al. 2021; Pezzuto et al. 2021) or for generating Purkinje trees (Barber et al. 2021; Berg et al. 2023; Biasi et al. 2025; Gillette, Matthias A. F. Gsell, et al. 2021). An additional challenge arises from the fact that some of these algorithms belong to closed pipelines, specifically tailored tools for a specific solver or purpose, sometimes even trained on highly specific data using machine learning algorithms, making their integration with other tools challenging.

In this study, we present an open-source computational pipeline designed to automatically generate all the data and files required to conduct simulation studies as needed for *in silico* clinical trials in patient-specific biventricular geometries. The pipeline focuses primarily on electrophysiological and electromechanical simulations, accounting for diverse pathologies and drug effects. We discuss each step of the method, with special attention to critical stages that may impact the automation process and their integration with existing algorithms. Furthermore, we demonstrate the application of the proposed methodology through a large cohort of geometries from the UK Biobank (UKBB) and other databases, showcasing its versatility and potential in effectively handling large databases.

# Methods

## Computational pipeline overview

We present a comprehensive pipeline for running large-scale simulation studies as in *in silico* trials, which encompasses the creation of all necessary information for the most common types of cardiac geometries acquired from magnetic resonance imaging (MRI) or computed tomography (CT). This pipeline, which takes as an input the patient anatomical surface mesh, operates on various types of biventricular geometries. It can automatically add labels for different cardiac structures, generate hexahedral or tetrahedral meshes for solving electrophysiological or electromechanical equations, and produce the most common fields required for these simulations, such as fibre information or electrophysiological (EP) heterogeneity. Additionally, the pipeline generates data fields compatible with new algorithms for Purkinje network generation (Berg et al. 2023), electrocardiogram (ECG) personalisation (Camps et al. 2024), and geometric data like Cobiveco coordinates (Schuler et al. 2021) or AHA segmentation (Cerqueira et al. 2002). In this work we used two different solvers for simulations: electromechanical simulations were run in Alya (Santiago et al. 2018, Wang et al. 2024) and electrophysiological simulations were performed using MonoAlg3D solver (Sachetto Oliveira et al. 2018).

The calculation of the generated fields, the geometry processing, and choice of solvers are guided by the design of the *in silico* pipeline illustrated in Figure 1. This figure highlights the pipeline's ability to handle a high number of simulations across large datasets, such as the UKBB. Depending on the purpose of the simulation, the generated fields are used to create various files that incorporate patient-specific variability in electrophysiology, drug dosages, or uncertainties in the simulation by adjusting parameters within defined ranges. Furthermore, additional variability or cardiac structures, such as the Purkinje networks, can be integrated into the simulations. The proposed pipeline supports these features, enhancing the scalability and adaptability of *in silico* cardiac trials.

In the following sections, we present the different stages of the framework and provide various examples of the pipeline's application and versatility. All the code used in this work is available at https://github.com/rdoste/InSilicoHeartGen. The UKBB meshes can be obtained through the UK Biobank database. Additionally, a virtual population of 100 synthetic meshes, based on the characteristics of the ones in the UKBB, was generated using a Variational Autoencoder (VAE) as described by (Beetz et al. 2023). These synthetic meshes replicate the morphological features of individual subjects while reflecting realistic population-wide diversity, similar to the original patient meshes. The pipeline can be tested on these virtual meshes, which are available for public access on Zenodo. Furthermore, other geometries used to validate the pipeline were obtained from open-access repositories, including datasets from studies such as (Pankewitz et al. 2024; Rodero et al. 2021; Schuler et al. 2021).

The pipeline presented here is implemented in MATLAB, with the exception of the mesh generation tools for which an interface with MATLAB is provided. It was tested on a laptop equipped with an Intel i7 processor, 16 GB of RAM, and an NVIDIA GTX 1650 GPU, as well as on other computers and laptops with similar specifications, demonstrating that it can be run efficiently on standard computational hardware.

**Design of *In silico* trials pipeline**

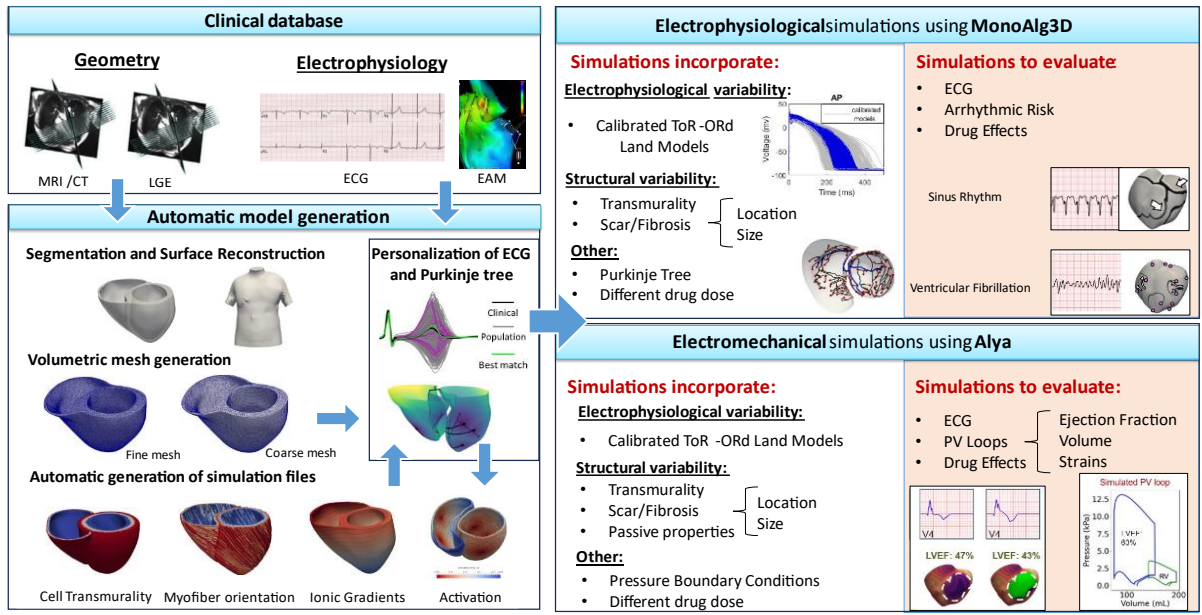

*Figure 1: Proposed workflow for large-scale modelling and simulation studies for in silico trials. Patient imaging data is used to automatically generate volumetric meshes and personalise them with the patient electrophysiological properties. The generated simulation files can be used for running electrophysiological or electromechanical simulations and can be adapted to various solvers. Depending on the final simulation purpose, additional variability or cardiac structures, such as the Purkinje network, can be integrated into the simulations. The presented pipeline supports all these features, enabling the automatic creation and execution of diverse simulation scenarios.*

1. Surface mesh generation

Image-based surface anatomical meshes are essential for generating volumetric meshes and assigning different labels with anatomical information. These meshes are created through the segmentation and reconstruction of MRI or CT images. Although this process may require some manual intervention, recent advancements have introduced algorithms capable of automating surface mesh generation in ventricles (Banerjee et al. 2021; Beetz et al. 2023; Beetz, Banerjee, and Grau 2022b, 2022c; Payer et al. 2018; Smith et al. 2024). However, semiautomatic segmentation remains necessary in most of the cases, particularly when extracting specific ventricular parts like outflow tracts, valves, or creating perfectly closed surface meshes to facilitate the volumetric mesh generation process. Nevertheless, the segmentation and surface mesh generation steps are heavily influenced by factors such as the centre where the images were acquired or the software utilised within the research group. Given the potential variability in these aspects, this study does not focus on this specific part of the pipeline and instead utilises the previous described algorithms to generate the surface meshes or works with previously generated meshes.

Despite the variability in segmentation and surface reconstruction, a closed surface mesh is always required to proceed with the following steps of the pipeline. This closed surface generation often involves closing gaps between surfaces, sealing the valves, or extruding the right ventricle (RV) surface, which is often comprised of a single surface. The proposed method for creating these meshes from the UKBB which are suitable for electromechanical simulation is described in the Supplementary Material. This algorithm automatically creates a closed mesh from the epicardial left ventricular (LV) surface and endocardial LV and RV surfaces. It generates an epicardial RV surface by extruding the RV endocardial surface by 3mm, merges the surfaces, and adds lids to the valve openings.

*Types of cardiac surface meshes*

Some factors, such as the segmentation method or the study's purpose, can produce various types of cardiac geometries. Typically, a full biventricular mesh, with information from the apex to the valves, is generated. Depending on the resolution of the acquisition and the image modality, some geometries can also present endocardial trabeculae.

In other cases, due to limitations in imaging or segmentation procedures, reconstructed cardiac geometries may only extend up to a truncated artificial plane placed below the valves, referred to in this work as "cut biventricular geometries".

Another type of biventricular geometry is one with no valve openings, often used for electromechanical simulations where pressure boundary conditions are applied at the endocardial valve surface. These closed geometries can be generated by either closing the valve openings in previously generated biventricular geometries or by directly generating the meshes from the images. This latter approach is used in this pipeline to automatically generate geometries for subjects of the UKBB dataset, employing multi-class point cloud completion networks (Beetz et al. 2023).

2. Surface Mesh Labelling

Mesh labelling is crucial for applying any algorithms and calculating various fields involved in the simulations. In most cases, mesh labelling can be performed using information from the image segmentation and reconstruction. However, in certain cases, this information is not included or it is incomplete. We present here an automatic algorithm for labelling cardiac geometries. The algorithm works for full biventricular geometries and for geometries with a cut artificial basal plane, the two most used types of geometries in research. A third version of the algorithm is also applied to UKBB geometries, used in this work. The method is able to automatically label the following surfaces in any given mesh: the basal plane, epicardium, RV and LV endocardium, RV septal endocardium, RV and LV apex and valve rings (if they are present).

The automatic label assignment is performed following a combination of the results of a ray tracing algorithm and several rules based on cardiac geometry. The algorithm generates rays from each mesh face centroid and calculates the number of intersections with other faces and the distance between them. A convex hull of the mesh is also calculated and included in the computation, facilitating the identification of epicardial faces and ensuring that every ray intersects with at least one face. For each ray, three different quantities are measured: number of ray intersections ($n_i$), distance of the closest face following origin face normal direction ($d_{n+}$), and distance of the closest face following opposite of face normal direction ($d_{n-}$). Since ray tracing algorithms can be computationally expensive, it is recommended to downsample the number of faces if the mesh contains several hundred thousand faces. The labels can be later projected back to a finer mesh without affecting the identification of structures.

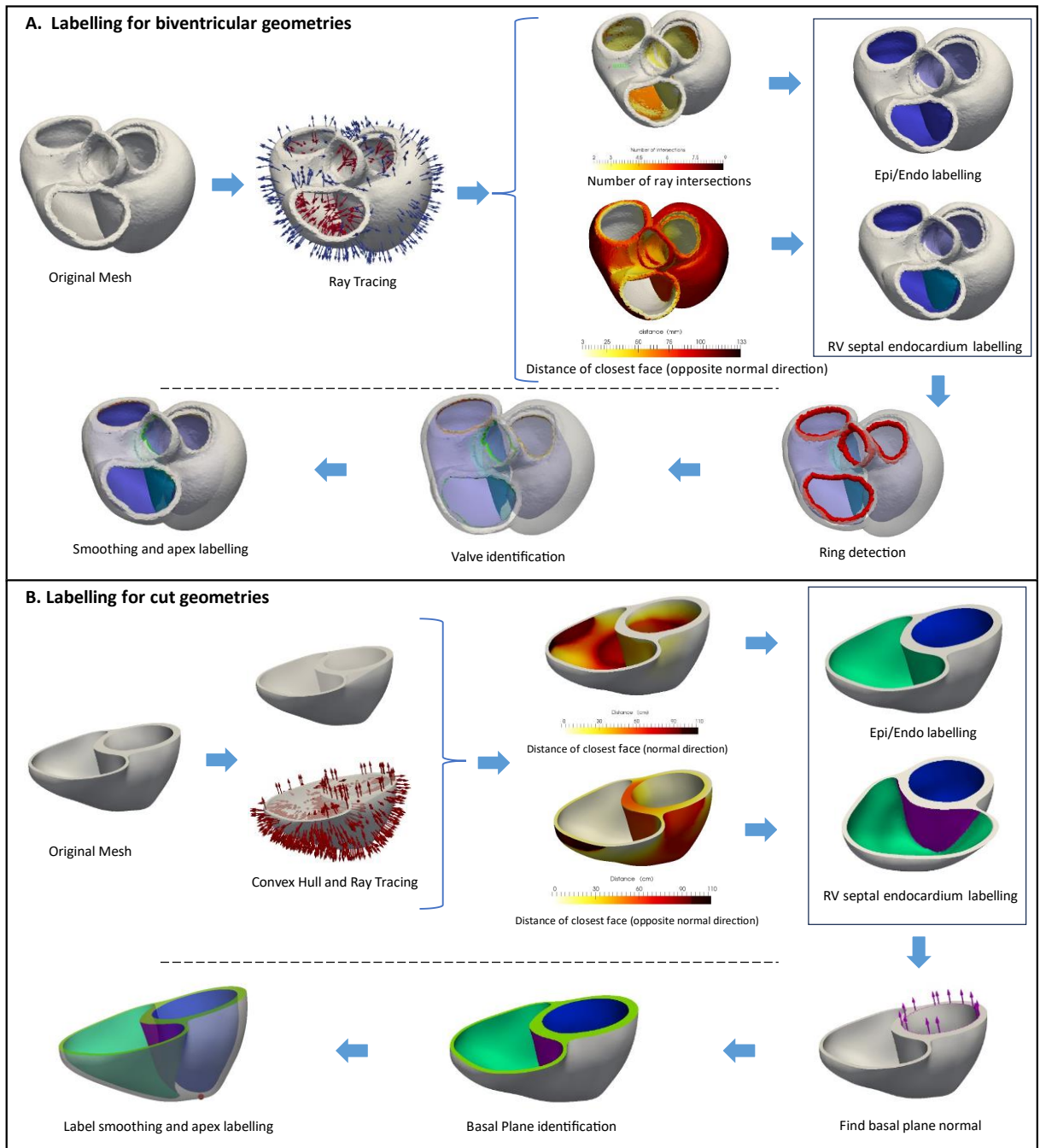

*Figure 2. Automatic labelling method for two different types of meshes: Biventricular geometry (A) and cut geometry (B). The method involves tracing rays from each face of the surface geometries. The number and distance of ray intersections are then calculated to determine the epicardium and endocardium for both the right and left ventricles, including the right septal endocardium. Once these surfaces are identified, their intersections are used to locate the valves in (A) or the basal plane in (B). Finally, the identified surfaces undergo smoothing to eliminate isolated patches, and the apices of the left and right ventricles are identified.*

*Description of the labelling algorithm*

*Biventricular geometry*

A description of this method can be found in Figure 2. Rays were traced from the centroid of each face, and the intersections with the rest of the mesh faces and the convex hull faces ($n_i$) were calculated.

This allows us to distinguish between the epicardium and the endocardial faces. The differentiation between the RV and LV can be achieved by computing the volume of the convex hull of each ventricle. In healthy conditions and in most pathologies, the RV volume is larger than the LV volume (Hudsmith† et al. 2005). Once identified, a smoothing algorithm is applied to reduce the number of detected surfaces to only three (epicardium, LV and RV). This smoothing algorithm considers the values of the neighbouring faces and adjusts them based on the majority value of neighbours.

After identifying these three surfaces, the subsequent steps involve identifying the RV septal endocardium. This surface is crucial for applying cell-type simulations in electrophysiology, since its electrophysiological properties resemble those of the LV epicardium (Konarzewska, Peeters, and Sanguinetti 1995). Moreover, some fibre generation algorithms (Piersanti et al. 2021) use this surface as an input. This identification is accomplished by utilising the distance of the closest face following the opposite direction of the face normal ($d_{n-}$). Since the thickness of the RV wall is usually smaller than the thickness in the septum, differentiation between the RV septal endocardium and the rest of the RV endocardium can be achieved by examining $d_{n-}$.

The final component of the labelling algorithm involves the identification of valves and apices. The location of the valves (ring detection) is easily computed by identifying the faces located in the borders between the epicardium and both endocardiums. Once RV and LV are detected, the remaining task is to identify the aortic versus mitral and pulmonary versus tricuspid valves. This is achieved by locating the centroid coordinates of each valve and calculating the distance of each valve relative to the others. The mitral valve is identified as the LV valve that is farthest from the RV valves. Meanwhile, RV valves are identified by calculating their distances relative to the mitral valve. Finally, LV and RV apices are determined by locating the long axes of the LV and RV ventricles, as described in (Schuler et al. 2021).

Electromechanical simulations are typically run on biventricular geometries with closed valves, which facilitates the calculation of pressures and results in more accurate deformation predictions. The method can be easily adapted to handle these geometries with some modifications. Closed valves need to be detected before running the labelling algorithm or can be added using rule-based methods (see Supplementary Material). Detection of the epicardial and endocardial surfaces can also be performed without the need for ray tracing, as these surfaces are independent and do not share any connectivity.

*Cut geometry*

Labelling in cut geometries follows a similar process, albeit with two main differences. Firstly, in this scenario, valve identification is omitted, and instead, the basal plane is automatically identified. Secondly, rays are traced from the face centroids, targeting only the faces of the convex hull (without calculating the intersections with the original mesh faces as in the biventricular geometry case).

In this case, the distance of the closest face following the face normal direction ($d_{n+}$) aids in identifying the epicardium. The optimal distance for distinguishing between epicardium and endocardium is determined iteratively, with a threshold value gradually increasing until two distinct clusters, corresponding to the RV and LV, can be discerned. The identification of RV versus LV endocardium and the septal RV endocardium follows a similar approach as in the biventricular case.

The second part of the algorithm involves identifying the basal plane. This is achieved by finding the normal of the basal plane faces. A practical and efficient approximation involves extracting the normal from the epicardial faces bordering the LV endocardium. Once the normal direction is extracted, all the epicardial faces that present an angle of less than 30 degrees are designated as the basal lid. A smoothing step is then applied to consolidate the lid labels into a single cluster.

Finally, the apex of the LV is located using a procedure similar to that employed in the biventricular case.

3. <u>Volumetric Mesh generation</u>

Electromechanical simulations and ECG personalisation computations are conducted on volumetric meshes, where equations are solved following spatial discretisation. The pipeline presented here demonstrates the capability to generate diverse mesh types at specified resolutions. In Figure 3, a schematic of the proposed meshing pipeline is depicted. For each geometry, two tetrahedral meshes are produced: one at a coarse resolution of 1500 µm edge length and one at a finer resolution (approximately 400 µm for only electrophysiological simulations and 1000 µm for electromechanical simulations). The coarser mesh is utilised for ECG personalisation and for accelerating the computing of the new fields, whereas the finer tetrahedral mesh, after value interpolation, is employed for precise simulations using the Finite Element Method on CPUs. Depending on the solver, the finer tetrahedral mesh can also be transformed into a hexahedral mesh to solve electrophysiology equations through the Finite Volume Method in GPU by employing the open-source software MonoAlg3D (Sachetto Oliveira et al. 2018). Interpolation between values at nodes of the coarse mesh and those of the fine or hexahedral mesh is performed using barycentric coordinates. For each node in the fine mesh, its position relative to the surrounding tetrahedron in the coarse mesh is determined. The values at the nodes of the coarse tetrahedron are then used to calculate the interpolated value at the new node.

All meshing tools employed in the pipeline are open source. Specifically, this study utilises the mmg3d

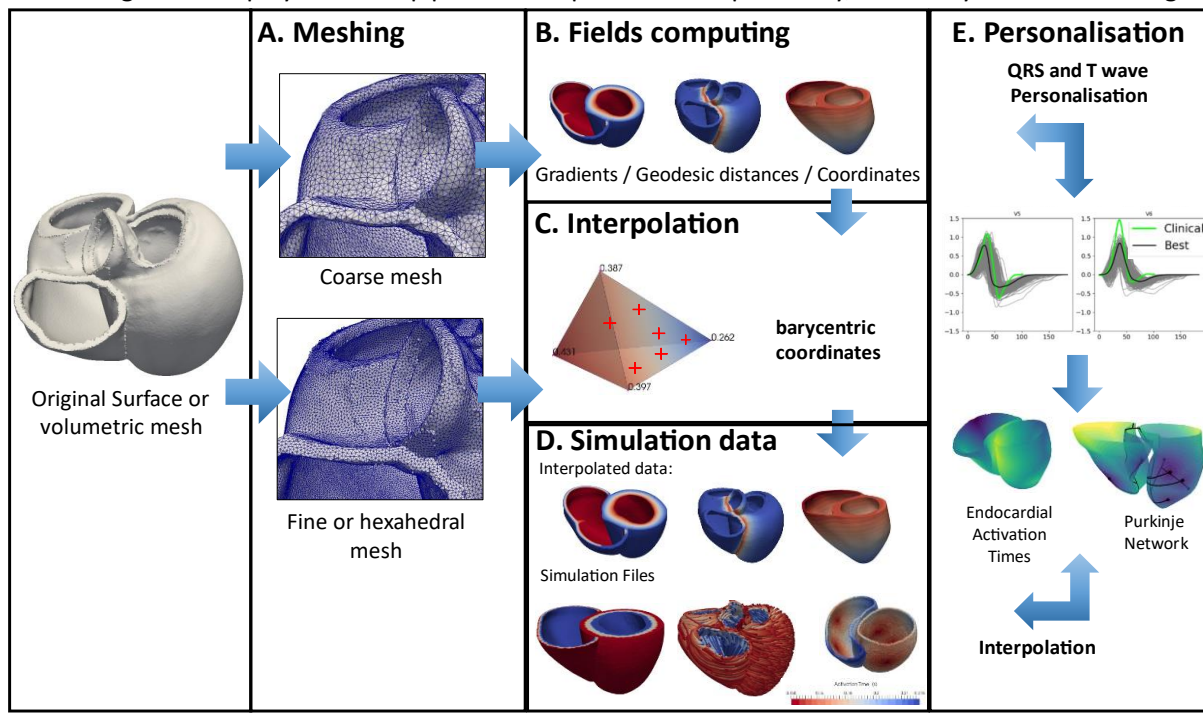

*Figure 3: Overview of the meshing and interpolation methodology. For each patient, a coarse and a fine/hexahedral mesh are generated (A). Coarse mesh is then used to calculate different gradients and coordinates (B). This fields can be also used for the electrophysiological personalisation or Purkinje generation (E) or to obtain the final simulations files (D), that are in the fine/hexahedral mesh after barycentric coordinate interpolation (C)*

meshing software(Dapogny, Dobrzynski, and Frey 2014) and Tetgen (through the Iso2mesh Matlab library (Tran, Yan, and Fang 2020)). The selection of these tools is driven by their versatility and diverse customisation options, enabling automated remeshing and definition of tetrahedral mesh edge lengths. For hexahedral meshes, a MATLAB function that is able to generate a vtk hexahedral mesh from a tetrahedral mesh is provided in the code (*hexa_mesher.m*). Other alternative meshing tools can also be easily integrated into the pipeline, thanks to its modular design.

Mesh quality was assessed by quantifying element distortions using the scaled Jacobian metric. This metric ranges from 0 to 1, where values closer to 1 indicate higher quality elements, and values near 0 represent lower quality.

### 4. Simulation files generation

Although each solver requires different types of files and formatting, the input data necessary for simulations can be automatically generated from specific fields, especially data that is highly dependent on geometry or cannot be directly personalised, such as fibre orientation, percentage of transmural cell types, or gradients in the expression of ionic channels. The method proposed in this study aims to standardise the processing methods applied to cardiac geometries, enabling the automatic generation of all required fields needed for simulations and post-processing of simulation results.

Figure 4 illustrates the most relevant fields, which are further described, along with some of their applications, in Table 1. All fields are initially calculated on the coarse mesh and then interpolated onto the finer or hexahedral mesh using tetrahedral barycentric coordinates. The fine mesh is where the different fields are processed and combined to prepare the information contained in the simulation files, such as the fibre orientation, AHA parcellation, cell type distribution, material properties or activation times.

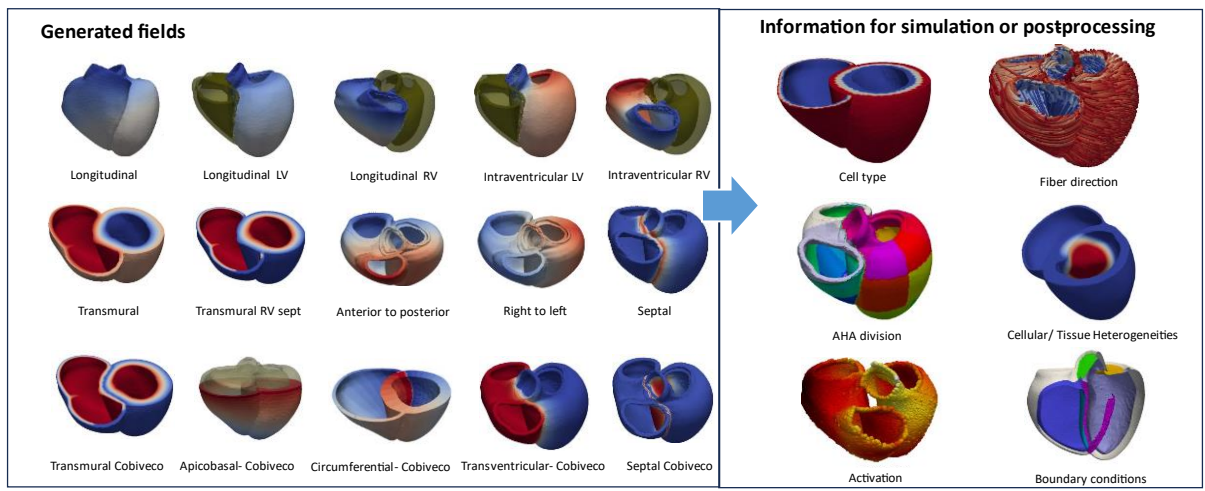

*Figure 4: Examples of the most common fields automatically generated by the pipeline (left), along with some of the information included in the simulation files (right), which are obtained by combining these fields.*

Table 1: Glossary of field labels for volumetric meshes generated by the automated pipeline and a list of how each field label is important for applications in the generation of various spatial fields and their significance in simulations.

| Field name | Description | Applications |
| --- | --- | --- |
| Longitudinal LV | Laplace solution from the LV apex to the LV valves | 1) LV apicobasal direction<br>2) Fiber generation (Doste et al. 2019) |

|  |  | 3) Define valve material |
| --- | --- | --- |
| Longitudinal RV | Laplace solution from the RV apex to the RV valves | 1) RV apicobasal direction<br>2) Fiber generation (Doste et al. 2019)<br>3) RV deformation computation (Bernardino et al. 2023) |
| Transmural | Distance from the endocardium to epicardium. Values in the septum consider LV as 2/3 of the total thickness. | 1) Fiber generation (Bayer et al. 2012; Doste et al. 2019)<br>2) Activation definition<br>3) AHA segmentation (Doste et al. 2019; Santiago et al. 2018)<br>4) Cell type Definition (Aguado-Sierra et al. 2022)<br>5) Purkinje network generation (Berg et al. 2023)<br>6) Trabeculae detection (Gonzalez-Martin et al. 2023) |
| Transmural RV | Similar to transmural, by defining the RV septal endocardial wall as epicardium | 1) Cell type definition<br>2) Fiber generation (Piersanti et al. 2021) |
| Intraventricular RV | Solution of the Laplace equation across the valves and apex of the RV ventricle | 1) Fiber generation (Doste et al. 2019)<br>2) Detection of outflow tracts (Doste et al. 2019) |
| Intraventricular LV | Solution of the Laplace equation across the valves and apex of the LV ventricle | 1) Fiber generation (Doste et al. 2019)<br>2) Detection of outflow tracts (Doste et al. 2019) |
| Longitudinal | Direction from the LV apex to all the valves | 1) Fiber generation (Bayer et al. 2012; Doste et al. 2019; Piersanti et al. 2021)<br>2) Apicobasal gradient |
| Septal | Distribution of values from the septum surface. It splits the septum as 2/3 LV and 1/3 RV. | 1) Fiber generation (Doste et al. 2019) |
| Left Ventricle to Right Ventricle | Direction from the LV to the RV | 1) Repolarisation personalisation (Camps et al. 2024; Gillette, Matthias A.F. Gsell, et al. 2021) |
| Posterior to anterior | Direction from posterior part of the ventricle to anterior | 1) Repolarisation personalisation (Camps et al. 2024; Gillette, Matthias A.F. Gsell, et al. 2021) |
| Septal-Cobiveco | Distribution of values in the septum as in Cobiveco coordinates | 1) Septum and Ridges definition (Schuler et al. 2021) |
| Apicobasal-Cobiveco | Ventricular Cobiveco coordinate | 1) Ventricular coordinates (Schuler et al. 2021)<br>2) Activation definition<br>3) Apicobasal gradient |
| Circumferential-Cobiveco | Ventricular Cobiveco coordinate | 1) Ventricular coordinates (Schuler et al. 2021)<br>2) AHA segmentation (Doste et al. 2019; Santiago et al. 2018) |

| Transmural-Cobiveco | Ventricular Cobiveco coordinate | 1) Ventricular coordinates (Schuler et al. 2021) |
| Transventricular-Cobiveco | Ventricular Cobiveco coordinate. Splits the septum in half. | 1) Ventricular coordinates (Schuler et al. 2021) |

*Field Computation*

The fields listed in Table 1 are computed using Laplace-Dirichlet Rule-Based Methods (LDRBMs), which solve the Laplace equation within the geometry while accounting for specific boundary conditions on the surfaces. These boundary surfaces are defined using the labels assigned during the labelling process, and distinct values are applied to these surfaces to solve the Laplace equation. The values and boundary conditions used for each field can be referenced in the works where they were first introduced. For the first eight fields in Table 1, their generation methodology is described in (Doste et al. 2019) while the fields associated with Cobiveco coordinates are detailed in (Schuler et al. 2021). The left-to-right ventricle and posterior-to-anterior values are calculated by normalising the projections of the nodal coordinates along their respective directions. The left-to-right direction is determined as the normal to the septal surface, while the posterior-to-anterior direction is defined as the cross-product of the septal surface normal and the apicobasal direction. Gradients extracted from these field values are also utilised to generate simulation inputs, such as those required for defining fibre orientations.

*Transference between complete biventricular geometries and cut biventricular geometries*

As discussed in the previous subsection, the pipeline supports both biventricular and cut biventricular geometries. However, certain algorithms, such as those used to compute Cobiveco coordinates, are specifically designed to operate on only one type of geometry, limiting their applicability to the other. This limitation is particularly evident in cases involving apicobasal coordinates or directions, where the definition of the basal plane can significantly affect results. To address this, we defined an artificial basal plane in the full biventricular geometries, located 1 cm below the pulmonary valve. This approach enables the accurate calculation of required fields with minimal distortion and facilitates the transfer of information between full and cut biventricular geometries.

*Ventricular coordinates*

Some of the proposed fields include ventricular coordinates. Defining a ventricular coordinate system allows for the parametrisation of cardiac geometry and transfer of data between different cases. The present method calculates biventricular coordinates using the Cobiveco methodology (Schuler et al. 2021) with some modifications. The framework performs this calculation without applying any remeshing, thereby avoiding the need to cut the mesh at basal plane and eliminating additional remeshing steps that could complicate the process and increase computation time. The septal and ridge surfaces, extracted after successive remeshings, are directly identified in the original mesh by locating the closest tetrahedral faces at those specific intersections. A comparison between the computed coordinates and original Cobiveco coordinates is provided in the Supplementary Material. Importantly, to remain consistent with the Cobiveco coordinate definition and avoid issues related to the bijective nature of the coordinates, no coordinates are assigned above the artificial basal plane.

*Fiber orientation*

In this framework, we used the algorithm published in (Doste et al. 2019) to calculate fibre orientation in biventricular geometries using the same proposed angles for left and right ventricles. The method relies on several fields defined in Table 1 to create a coordinate system that determines fibre orientation at each node of the mesh. Additionally, the pipeline allows users to modify these angles, enabling adjustments to the fibre orientation by any specified value.

*Adding variability in the simulation files*

*In silico* trials simulations of cardiac activity must account for variability among patients. This includes changes in the characteristics of the action potential, tissue passive properties, pressure boundary conditions and other factors. With the proposed automatic pipeline, multiple cases can be created simultaneously, significantly reducing processing time. An important aspect is incorporating variability at the cellular level. As done in previous studies (Coleman et al. 2024; Dasí et al. 2022) , this is achieved by introducing variability in ionic current densities through different scaling factors. A similar approach can be applied to controlled inputs, such as drug dosage (Camps et al. 2025).

5. Electrophysiology and conduction velocities personalisation

Electrophysiological personalisation is essential to obtain patient-specific models. This personalisation is often done by obtaining an activation pattern that can reproduce the patient ECG. Recently, several works have developed new algorithms to personalise the ventricle EP (Camps et al. 2024; Gillette, Matthias A.F. Gsell, et al. 2021).

The presented pipeline is designed to support personalisation by including all the necessary fields and gradients required to execute the personalisation. In cases where no electrophysiology personalisation is needed, the pipeline projects the activation of several endocardial nodes from a specific geometry to the endocardium of each new geometry using the Cobiveco coordinates. Electrode locations are also automatically generated by registering the heart geometries using three main hearts directions as defined in (Schuler et al. 2021). Additionally, the pipeline is capable of incorporating personalised torso geometries extracted automatically (Smith et al. 2022) and personalised electrode placement (Li et al. 2025).

Finally, the automatic fields obtained by the method can be used to generate patient-specific Purkinje networks using open source code by providing the endocardium surfaces from the LV and RV (Berg et al. 2023).

6. Applications

To demonstrate its capability and functionality, the proposed pipeline was used in three applications: 1) to generate files for electromechanical simulations on a cohort of 100 patients from the UK Biobank, showcasing its potential to handle large-scale databases and perform examples of *in silico* trials on a small group of patients to demonstrate how the pipeline can be used to plan studies investigating the effects of drugs using electromechanical simulations, 2) generation of four EP digital twins, incorporating personalised electrophysiology and torso geometries, and 3) the generation of 20 hexahedral meshes using the Rodero dataset for electrophysiological simulations using the open-source EP solver MonoAlg3D (Sachetto Oliveira et al. 2018). The aim of these simulations is to demonstrate the versatility of the open-source pipeline for integrating all necessary steps to facilitate electrophysiological and electromechanical simulations for *in silico* trials.

*6.1 Electromechanical Simulations of 100 subjects from the UK Biobank*

Ventricular electromechanical simulation were conducted using the high-performance numerical software Alya (Santiago et al. 2018) on the ARCHER2 supercomputing centre. The pipeline developed in this study was tested to generate biventricular models from the MRI data of healthy patients from the UK Biobank, create all the necessary files for the simulations, and perform multiscale simulations that range from cellular ionic currents to body surface ECG and pressure-volume loops.

The electrical propagation was modelled using the monodomain equation with orthotropic diffusion based on fibre direction. Tissue diffusivities were calibrated to match experimentally measured orthotropic conduction velocities of 67 cm/s, 30 cm/s, and 17 cm/s (Caldwell et al. 2009) while transmural and apex-to-base heterogeneities (Mincholé et al. 2019) were incorporated to replicate experimentally reported gradients (Chauhan et al. 2006). Electrical activation during sinus rhythm was simulated via a fast-activation layer along the endocardium, mimicking propagation through the subendocardial Purkinje network and achieving realistic simulated QRS complex morphologies. Activation patterns were standardised between geometries by transferring local activation times using Cobiveco coordinates.

Human ventricular membrane kinetics were represented by the ToR-ORd model (Tomek et al. 2019) given its demonstrated validity across healthy, diseased, and drug-block conditions. This model was coupled with human excitation-contraction and active tension Land model (Land et al. 2017). For each patient, a population of nine different cardiomyocyte models was created by introducing variability in the ionic current densities of the cellular model. These models were extracted from a previously calibrated control population (Doste, Coppini, and Bueno-Orovio 2022) according to the ranges extracted from human experimental data. At organ and tissue levels, the model incorporated strongly-coupled electromechanics, which included orthotropic passive mechanical behaviour and the balance of linear momentum with inertia effects (Levrero-Florencio et al. 2020). A pericardial constraint was simulated by applying an elastic spring boundary condition perpendicular to the epicardial surface. Pressure boundary conditions, as well as active tension and stiffness parameters, were also calibrated as in (Wang et al. 2021; Zhou et al. 2022).

For each case, simulations were conducted for 2 seconds with a cycle length of 800 ms, during which the effects of Verapamil on the ECG, pressure, and volume were evaluated. Two different doses of Verapamil (0.1 μM and 0.2 μM) were used.

*6.2 Building Digital Twins: Personalisation of Electrophysiology and Patient Torso of 4 subjects*

In this second application, heart and torso geometries from UKBB patients were used to build digital twins (Corral-Acero et al. 2020) generating the simulation files and personalising the electrophysiology to match the patient ECG. Two types of geometries from the UKBB were employed for personalisation: the previously generated geometries and cut biventricular geometries, which were obtained through a different segmentation procedure as described in (Banerjee et al. 2021). Personalisation of the patient-specific activation and tissue conduction velocities was conducted using the algorithm presented by (Camps et al. 2024). Once the activation data was obtained, it was exported to the simulation files to ensure that the EM simulation would accurately reproduce the activation pattern. Due to the high computational demands of electrophysiological personalisation, this process was applied to four different patients.

*6.3 GPU-Enabled Electrophysiology Simulations of 20 subjects from the Rodero dataset*

For the third application of the pipeline, a different database and simulation software were employed to conduct electrophysiology simulations. A set of 20 biventricular geometries, derived from the study by (Rodero et al. 2021), was used. Electrophysiology simulations were conducted using MonoAlg3D, an open-source, GPU-enabled solver. The automatic pipeline generated hexahedral meshes with a 400 μm edge length, applying the automatic labelling algorithm to the surface meshes to ensure all necessary tags were accurately assigned, despite some labels being initially available. After creating the required fields on the coarse mesh, they were interpolated onto the hexahedral mesh. Electrical propagation was simulated using the monodomain equation, and cellular membrane kinetics were also represented by the ToR-ORd model for human ventricular cells.

7. <u>Test meshes</u>

To facilitate the application of this pipeline and enable the development of new simulations and algorithms by research groups without direct access to the UKBB data, we generated a virtual population of meshes using a Variational Autoencoder (VAE) as described by (Beetz et al. 2023; Beetz, Banerjee, and Grau 2022a). This method produces new geometries that retain similar shapes and subpopulation-specific variations, consistent with the original dataset. The generated cohort of meshes, including all the generated fields, is made available for download to support further research and application development. The format of the meshes and the outputs used and produced by the pipeline is described in the Supplementary Material.

## Results

**Labelling results**

Since the pipeline is automated, assigning proper input labels to the closed mesh is a crucial step that directly affects the results of the simulation. Therefore, an analysis of the labelling is performed. Figure 5a shows the results of applying the automatic labelling code to different types of heart geometries, demonstrating its versatility in handling open and closed biventricular geometries. In cases of closed geometries, the code is also capable of including an approximate valve location.

A quantitative evaluation of the heart labelling performance is not straightforward and requires substantial manual work to establish an acceptable "ground truth" to compare our result. To address this challenge, we demonstrate the robustness of the presented algorithm by comparing the mean distance of several labels generated by our method, using an unlabelled mesh as an input, against the labels provided in an open-source dataset created using segmentation information. We used 20 biventricular geometries obtained from CT scans by (Rodero et al. 2021). In addition, some transformations were made using the available data (ID identification and RHO values) to obtain similar tags as the ones provided by our algorithm, and valve location was obtained calculating the intersection between the valves and the biventricular mesh. Figure 5b shows two meshes for a specific subject: one with labels from the dataset (left) and the mesh with labels calculated with our algorithm (right). Figure 6b also shows the mean distance between faces of the epicardium, endocardium of the left and right ventricles, septal right ventricle, and valves as obtained by our algorithm, compared to the faces of reference meshes. The measured mean distances were low (less than 2 mm), indicating a good overlap between the detected surfaces in labelling process and those in the reference meshes.

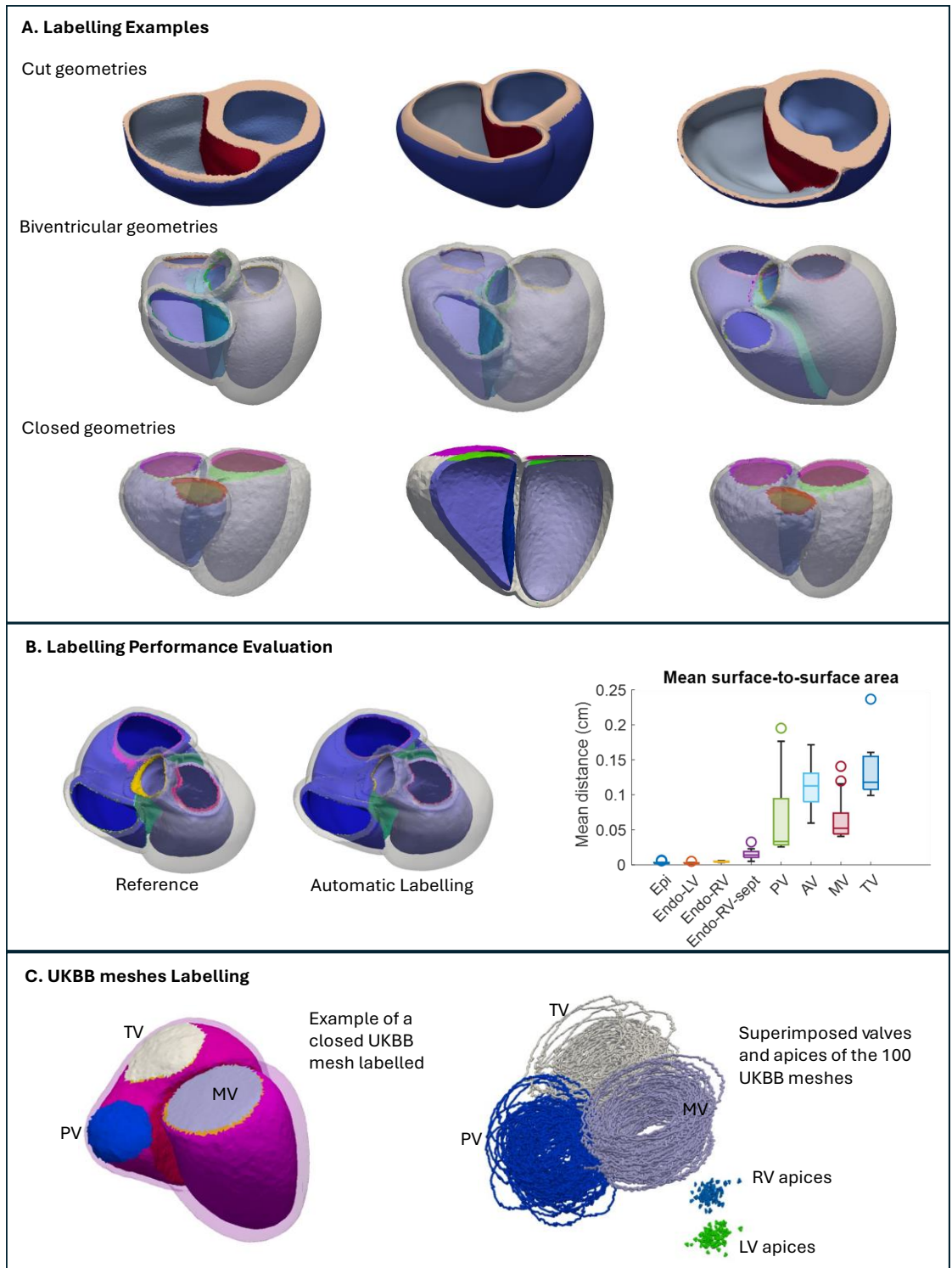

*Figure 5: A) Resulting labels applied in different heart geometries. B) Automatic labelling method applied to 20 meshes, and results compared with pre-existing segmentation data. The mean distance between corresponding faces was measured for each labelled element (right), with lower values indicating a good correspondence. In (C), the labelling method was applied to 100 meshes from the UK Biobank (UKBB) dataset, which were centered and aligned. A sample case is shown at the left, while the contours of the valves and apices are displayed at the right. No misplaced or abnormally labelled valves or apices were observed.*

The precision and accuracy from the algorithm in creating the valves in the UKBB closed geometries is shown in Figure 5c, where the different valves created in each of the 100 UKBB geometries are superimposed, including the RV and LV apices. This representation allows us to confirm that no valves or apices are misplaced during the labelling process.

**Workflow performance and mesh quality**

The tetrahedral meshes generated in this study presented an average scaled Jacobian of 0.65 , with values ranging from 0.13 to 0.99. These results ensure that the meshes are of sufficient quality to support electromechanical simulations without compromising numerical accuracy.

Mesh resolution significantly affects the performance of the workflow. In this work, a resolution of 1 mm was selected for electromechanical meshes, while a finer resolution of 0.4 mm was used for electrophysiological simulations. The average processing time per case—including the generation of a closed surface mesh, labelling, volumetric (tetrahedral and hexahedral) mesh generation, and simulation file preparation—was approximately 10 minutes.

The pipeline was able to run automatically for all cases that had a pre-existing closed surface mesh. Only 3 out of the 100 UKBB meshes (3%) required any manual intervention. These exceptions occurred during the step of creating a closed mesh, where the mesh repair algorithms were unable to resolve issues directly and required manual corrections.

**Pipeline applications: evaluating drug response variability, personalising electrophysiology and demonstrating its versatility.**

A proof of concept demonstrating the utility of this pipeline in electromechanical *in silico* trials was conducted by simulating the effects of control, 0.1 μM, and 0.2 μM doses of Verapamil on different patients, incorporating EP variability through the use of 9 distinct cellular models. All required simulation files were generated using the pipeline, resulting in a total of 27 cases per patient. These cases were simulated in an HPC facility, with each case taking approximately 2 hours of real time to complete.

Figure 6 presents the ECG and pressure-volume (PV) loops obtained for two different patients. The simulations yielded results consistent with clinical data (Nguyen et al. 2017; Vicente et al. 2015), such as a decrease in T-wave amplitude and a reduction in contractility, primarily due to the reduction of the L-type calcium current. It is essential to highlight that the variability in the *in silico* trials can substantially impact the outcomes. Notably, each cellular model exhibited distinct QT intervals, and the degree of change in the PV loop induced by Verapamil varied across the distinct cellular models. Additionally, as expected, the personalisation of the fields to patient-specific geometries greatly influenced the PV loops. For instance, Patient A exhibited a larger volume and reached higher pressures. This example underscores the importance of incorporating variability in drug-related *in silico* trials, as such variability can significantly affect the results.

The proposed pipeline was used to generate the gradients and electrode positions required to obtain personalised activation for several subjects from the UKBB, including different types of biventricular geometries. Figure 7A shows the results of 3 cases obtained after adapting the pipeline to work with an existing algorithm (Camps et al. 2024). The left column presents the local activation times, which serve as input for the *in silico* trial geometries, while the right column displays a comparison between the clinical and inferred QRS complexes.

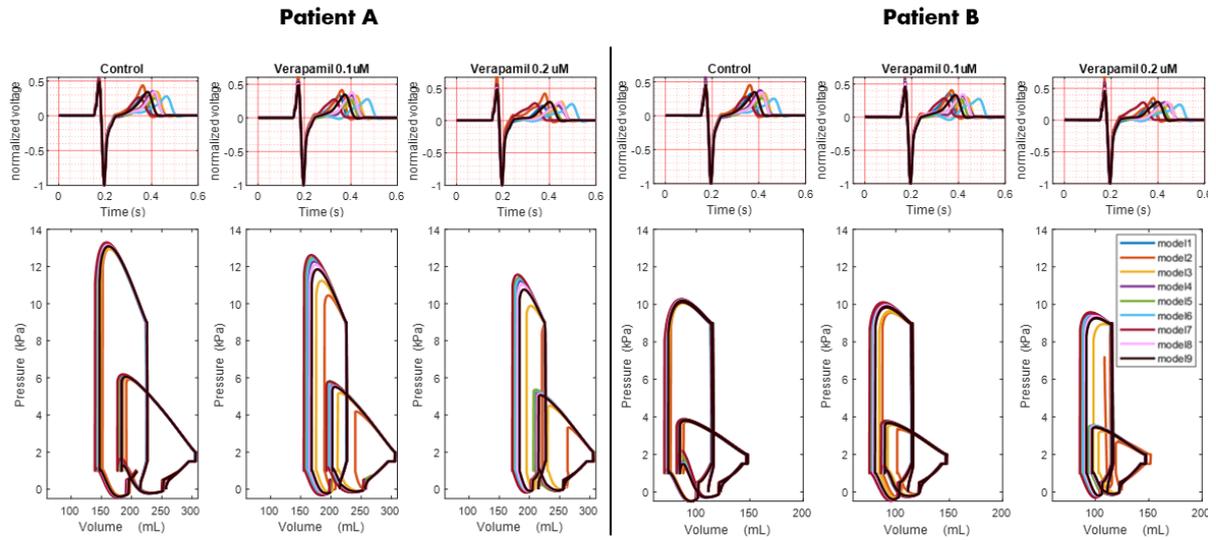

*Figure 6: ECG and PV loops of two different patients simulated using* in silico *drug trials for control, 0.1 μM, and 0.2 μM doses of Verapamil. While most of the models showed a decrease in T wave amplitude and a reduction in contractility in the PV loop, the different models exhibited varying responses to the drug doses.*

The final application of the proposed pipeline was to adapt it for use with a fully open-source EP solver. In this case, we employed the MonoAlg3D solver, which required all data to be interpolated onto hexahedral meshes. Figure 7B illustrates three different types of hexahedral geometries used for EP simulations. Additionally, the pipeline was made compatible with a Purkinje network generation method (Berg et al. 2023), as it can be observed in the Figure 7B. Results from the simulations can be observed in Figure 7B (bottom row), which shows the ECGs obtained from several of the 20 CT-based geometries processed by the pipeline while maintaining the same activation pattern.

## A. Activation Personalisation

**Activation Time Maps**   **ECG**

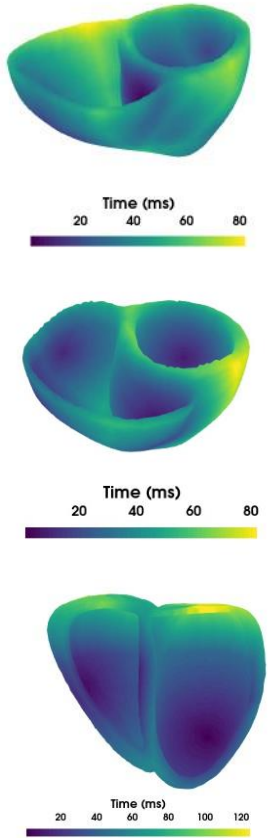
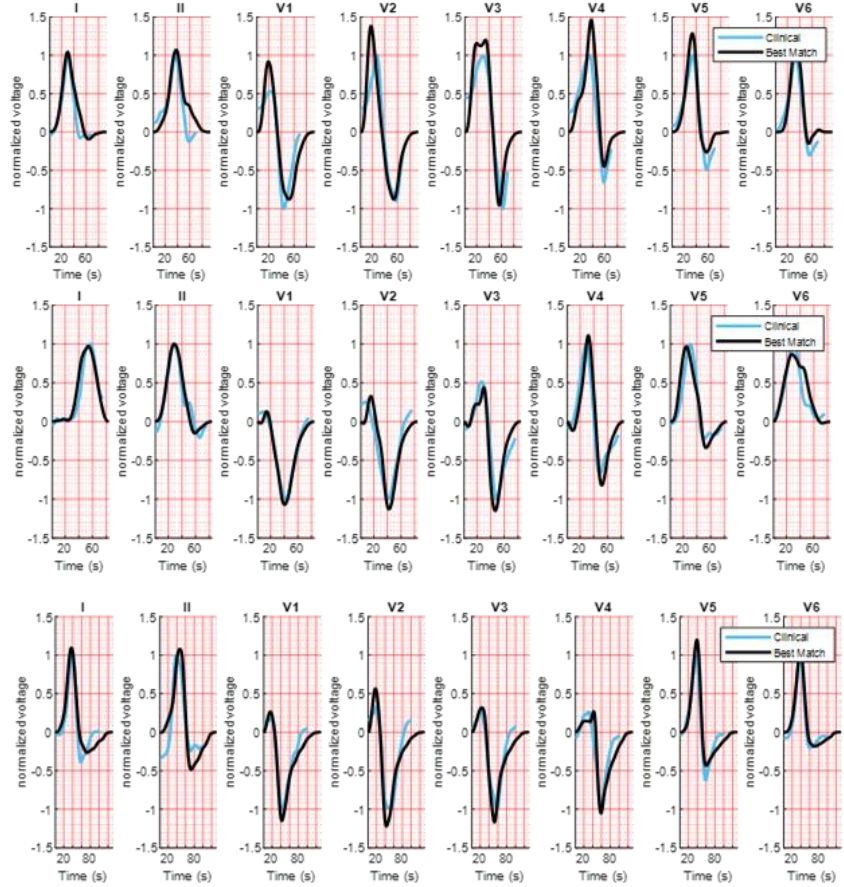

## B. Hexahedral Meshes for GPU simulations

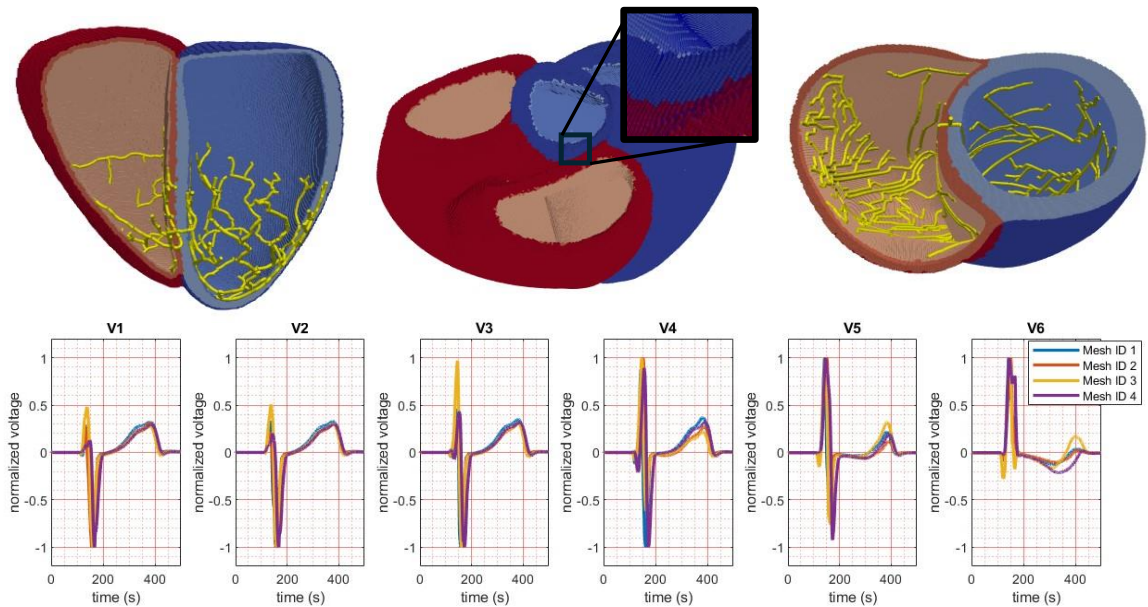

*Figure7. (A) Different types of hexahedral meshes created to run for MonoAlg3D. Two of them are plotted together with a personalised Purkinje network. (B) ECGs obtained from simulating sinus rhythm in geometries obtained from different datasets.*

# Discussion

In this work, we present a pipeline for automating *in silico* trials across large cohorts of human ventricular geometries. The proposed pipeline enables researchers to automatically create all the files needed for electrophysiological or electromechanical biventricular simulations, while also allowing the personalisation of key parameters within a digital twin framework and addressing both electrophysiological and structural variability. Additionally, the code is open-source, enhancing the pipeline's versatility and reproducibility.

The proposed pipeline introduces several key innovations, not included in previous works, that make it suitable for *in silico* trials:

- **Automatic generation of simulation input files:** The fully automated workflow generates all the necessary inputs for running electromechanical and electrophysiological simulations. By automating this process, the pipeline reduces human intervention and error, improving the efficiency in large-scale simulations.
- **New automatic heart labelling algorithm:** We developed a novel heart labelling algorithm capable of handling the two most common heart geometries used in EP simulations. This facilitates the application of this pipeline and other existing algorithms to a variety of geometrical representations of the heart.
- **Algorithm integration**: The pipeline has integrated several algorithms for fibre orientation generation, ventricular coordinates generation, AHA segmentation or cell type generation. This allows users to easily generate new simulation cases by simply adjusting or customising key parameters. This flexibility makes it adaptable to different study designs, patient-specific models, or drug testing scenarios.
- **Efficient use of meshes for personalisation:** The pipeline is capable of simultaneously processing both fine and coarse meshes. This is particularly advantageous for integration with digital twin pipelines that use fast simulation methods, such as Eikonal or phenomenological models, to adjust earliest activation nodes or Purkinje networks. This ability to efficiently handle different mesh resolutions reduces processing time without compromising the accuracy of the simulation outcomes.

Automatic heart labelling

The labelling method in the proposed pipeline is rule-based, providing a higher level of flexibility when dealing with new or previously unseen geometries. Since this method does not require extensive training for labelling, it offers significant advantages over machine learning-based approaches, particularly in terms of processing time and generalisation to new datasets.

Accurate labelling is crucial to ensure reliable simulation results. Our evaluation demonstrated that the labelling algorithm effectively handles the most commonly used biventricular geometries in cardiac simulations, yielding results with minimal differences compared to methods that assign labels during image segmentation. While label assignment during segmentation can be more optimal, it is not always feasible. The proposed labelling algorithm addresses this limitation, enabling users to standardise inputs and overcome inconsistencies in geometry preparation. It also allows users to reuse previous geometries from different studies by easily incorporating any missing fields or information not included in the original mesh. Additionally, it can automate the generation of input data for algorithms that previously required manual intervention, such as those presented in (Biasi et al. 2025; Pankewitz et al. 2024; Piersanti et al. 2021).

Design of the pipeline

The presented pipeline was designed according to the planning outlined in Figure 1. All calculated fields have been shown in previous works to play a critical role in *in silico* trials, enabling the evaluation of ECGs, arrhythmic risk, PV loops, and drug effects (Dasí et al. 2024; Gonzalez-Martin et al. 2023; Martinez-navarro et al. 2020; Zhou et al. 2022). The open-source and flexibility nature of the pipeline ensures that any required information not initially included can be incorporated in the future. This includes personalisation information such as scar distribution, fibrosis patterns, or electro-anatomical mapping data, which are not available in the UKBB dataset. Furthermore, the pipeline's modularity allows for the use of alternative mesh resolutions or the addition of fields beyond those described in Section 2.4 at any stage of the process.

The pipeline processes each case in approximately 10 minutes, which is remarkably fast given the complexity of cardiac simulations. This processing time enables the pipeline to handle the large-scale demands of datasets such as the UKBB or pharmaceutical trials, where thousands of cases may need to be simulated under various conditions.

We tested the pipeline on 100 cases from the UKBB and 100 synthetic meshes, with only 3% of cases requiring additional manual intervention, demonstrating the robustness of the approach. The synthetic meshes, along with all the generated files, are publicly available on ZENODO, allowing for visualisation and further validation.

Electromechanical *in silico* drug trials were conducted on a subset of patients to demonstrate the pipeline's ability to enable such applications and generate meshes capable of reproducing valid ECGs and PV loops. The results presented in Figure 7 underscore the significant impact of anatomical and electrophysiological variability on outcomes, emphasising the importance of optimal patient stratification in determining which patients are most likely to respond favourably to a drug. This aligns with findings from prior *in silico* drug trials on atrial geometries (Dasí et al. 2024). These findings further highlight the importance of using an automated pipeline to prepare all required fields and information for simulations, ensuring that such variability is accurately accounted for.

Pipeline's versatility and integration with other algorithms

Unlike solver-oriented environments, where tools are specifically designed for use with electrophysiological or electromechanical solver, the pipeline presented here is highly versatile and can be integrated with a variety of open-source tools. As shown in Figure 8, the pipeline integrates with personalisation algorithms, enabling the customisation of conduction velocities, depolarisation, and even repolarisation of the patient's ECG. The pipeline can also create inputs for the generation of a Purkinje networks, facilitating a more detailed study of EP mechanisms and arrhythmic risk in patients when the cardiac conduction system is considered. Overall, this pipeline offers the capability to merge these open-source algorithms and streamline the process from patient imaging to electromechanical and electrophysiological simulations of digital twins. This is especially significant when running EP simulations using MonoAlg3D, a purely open-source EP solver, which allows researchers to generate cases directly from imaging to patient-specific geometries and incorporate all necessary simulation data. The solver's compatibility with Purkinje networks' structures further enhances the pipeline's value by integrating cutting-edge technology.

The pipeline's versatility has been demonstrated through its application to different patient geometries and datasets. The fields generated by the pipeline are also compatible with other algorithms, as shown in Table 1. In this work, we chose Cobiveco coordinates to describe patient geometry, which, as highlighted in the original paper (Schuler et al. 2021) present several advantages over their predecessors, the UVC coordinates (Bayer et al. 2018), including better symmetry, consistency, and bijectivity. Although this method results in the loss of information in the basal part of the heart, even for biventricular geometries it is more efficient for transferring data across majority of the ventricle. Recently, CobivecoX coordinates (Pankewitz et al. 2024) have been introduced to address this issue; however, this approach could not be implemented on UKBB meshes due to missing aortic valve imaging information (Banerjee et al. 2021).

Limitations:

The pipeline has the following limitations:

- The pipeline operates automatically, which, while efficient, may lead to occasional mislabelling or improper detection of certain labels, especially when a new or atypical geometry is processed. Such issues are more likely to arise when dealing with abnormal patient geometries caused by certain pathologies. However, the code includes built-in methods for detecting these potential errors and alerting the user. Additionally, the pipeline provides several parameters that can be adjusted to fine-tune the results.

- The code has been tested in local computing environments. Implementation on high-performance computing systems could significantly accelerate processing times by reducing data transfer overhead and enabling parallel computation

- The code works from surface meshes, meaning that external tools are required for segmentation and surface mesh reconstruction.

- The implementation of ventricular coordinates on the basal part of the heart is not straightforward due to the complex shape and lack of symmetry, especially in the right ventricle. While some approaches, such as CobivecoX, have been proposed, their implementation is not always feasible and depends on the specific geometry being used. Our pipeline currently only allows data transfer below the basal plane, which is a valid approach for most applications. However, this method may not be suitable in certain cases, such as when outflow tracts are involved.

- The pipeline is primarily designed for use with data from the UKBB, meaning that certain types of information, such as scar location, fibrosis, and electroanatomical maps, are not included by default. However, due to the open-source nature of the pipeline, these additional data types can be incorporated with relative ease, allowing for greater customisation and expansion in future work.

# Conclusions

In this work, we present open-source, automatic pipeline for generating and processing patient-specific models on a large scale. The pipeline labels the mesh to create necessary simulation files and allows for model personalisation when required. By utilising coarse meshes for field generation and electrophysiology personalisation, it enables efficient preparation of simulation cases. This speed, combined with automation, is critical for *in silico* trials, where uncertainties and variability—such as clinical protocols, drug dosages, or electrophysiological variability—demand a large number of simulations. This aspect highlights the pipeline's potential to support pharmaceutical research and drug development through *in silico* trials.

Acknowledgements

This work was funded by a Wellcome Trust Fellowship in Basic Biomedical Sciences to Blanca Rodriguez (214290/Z/18/Z) and the CompBioMed Centre of Excellence in Computational Biomedicine (European Commission Horizon 2020 research and innovation programme, grant agreement No. 823712). Abhirup Banerjee is supported by the Royal Society University Research Fellowship (Grant No. URF\R1\221314). The computation costs were incurred through the CompBioMedX project (EP/X019446/1), which provided access to ARCHER2, a UKRI national supercomputing service. This study also used high-performance computing resources from the Polaris supercomputer at the Argonne Leadership Computing Facility (ALCF), Argonne National Laboratory, United States of America. The U.S. Department of Energy's (DOE) Innovative and Novel Computational Impact on Theory and Experiment (INCITE) Program awarded access to Polaris. The ACLF is supported by the Office of Science of the U.S. DOE under Contract No. DE-AC02-06CH11357. For the purpose of Open Access, the authors have applied a CC BY public copyright licence to any Author Accepted Manuscript (AAM) version arising from this submission.